\documentstyle[12pt]{article}
\begin{document} 
\vspace*{-1in} 
\renewcommand{\thefootnote}{\fnsymbol{footnote}} 
\begin{flushright} 
TIFR/TH/99-13\\
TIFR-HECR-99-03\\
April 1999\\ 
\end{flushright} 
\vskip 65pt 
\begin{center} 
{\Large \bf Testing TeV Scale Quantum Gravity Using Dijet Production
at the Tevatron}\\
\vspace{8mm} 
{\bf 
Prakash Mathews${}^{1}$\footnote{prakash@theory.tifr.res.in}, 
Sreerup Raychaudhuri${}^2$\footnote{sreerup@iris.hecr.tifr.res.in},   
K.~Sridhar${}^1$\footnote{sridhar@theory.tifr.res.in}
}\\ 
\vspace{10pt} 
{\sf 1) Department of Theoretical Physics, Tata Institute of 
Fundamental Research,\\  
Homi Bhabha Road, Bombay 400 005, India. 

2) Department of High Energy Physics, 
Tata Institute of Fundamental Research, \\  
Homi Bhabha Road, Bombay 400 005, India. 
\footnote{Address after May 1, 1998~: Department of Physics, 
Indian Institute of Technology,\\ 
Kanpur 208 016, India.}} 
 
\vspace{80pt} 
{\bf ABSTRACT} 
\end{center} 
\vskip12pt 
\noindent Dijet production at the Tevatron including effects of
virtual exchanges of spin-2 Kaluza-Klein modes in theories
with large extra dimensions is considered. The experimental
dijet mass and angular distribution are exploited to obtain
stringent limits ($\sim 1.2$ TeV) on the effective string scale 
$M_S$.
\setcounter{footnote}{0} 
\renewcommand{\thefootnote}{\arabic{footnote}} 
 
\vfill 
\clearpage 
\setcounter{page}{1} 
\pagestyle{plain}
\noindent There have recently been major breakthroughs in the understanding of
string theories at strong coupling in the framework of what is now known as 
$M$-theory \cite{hw, polchinski, lykken}. In particular, unification of gravity 
with other interactions now seems possible in the $M$-theoretic 
framework. But of tremendous interest to phenomenology
is the possibility that the effects of gravity could become large at
very low scales ($\sim$~TeV), because of the effects of large extra
compact dimensions where gravity can propagate \cite{dimo}. 
Starting from a higher-dimensional theory of open and closed strings 
\cite{dimo2, shiu}, the effective low-energy theory is obtained by
compactifying to
3+1 dimensions, in such a way that $n$ of these extra dimensions
are compactified to a common scale $R$ which is large,
while the remaining dimensions are compactified to 
extremely tiny scales which are of the order of the inverse Planck scale. 
In such a scenario, the Standard Model (SM) particles correspond to 
open strings, which end on a 3-brane and are, therefore, confined to 
the $(3+1)$-dimensional spacetime. 
On the other hand, the gravitons (corresponding to closed strings) propagate 
in the $(4+n)$-dimensional bulk. The relation between 
the scales in $(4+n)$ dimensions and in $4$ dimensions is given by \cite{dimo}
\begin{equation} 
M^2_{\rm P}=M_{S}^{n+2} R^n ~,
\label{e1} 
\end{equation} 
where $M_S$ is the low-energy effective string scale. This equation has
the interesting consequence that we can choose $M_S$ to be of the order
of a TeV and thus get around the hierarchy problem. For such a value of
$M_S$, it follows that $R=10^{32/n -19}$~m, and so we find that $M_S$
can be arranged to be a TeV for any value $n > 1$. Effects of non-Newtonian
gravity can become apparent at these surprisingly low values of energy.
For example, for $n=2$ the compactified dimensions
are of the order of 1 mm, just below the experimentally tested region
for the validity of Newton's law of gravitation and within the possible
reach of ongoing experiments \cite{gravexp}.
In fact, it has been shown \cite{dimo4} that is 
possible to construct a phenomenologically viable scenario with large
extra dimensions, which can survive the existing astrophysical and 
cosmological constraints.
For some early papers on large Kaluza-Klein dimensions, see Ref.~\cite{anto, 
taylor} and for recent investigations on different aspects of the
TeV scale quantum gravity scenario and related ideas, see Ref.~\cite{related}.

Below the scale $M_S$ \cite{sundrum, grw, hlz}, we have an
effective theory with an infinite tower of massive Kaluza-Klein
states. 
which contain spin-2, spin-1 and spin-0 excitations. The spin-1
couplings to the SM particles in the low-energy effective
theory are not important, whereas the scalar modes couple to the trace
of the energy-momentum tensor, which vanishes for massless
particles. Other particles related to brane dynamics 
(for example, the $Y$ modes which are related to the
deformation of the brane) have effects which are subleading, compared to
those of the graviton. The only states, then, that contribute 
to low-energy phenomenology are the spin-2 Kaluza-Klein states. 
For graviton momenta smaller than the scale $M_S$, the
effective description reduces to one where the gravitons in the bulk 
propagate in the flat background and couple to the SM fields 
via a (four-dimensional) induced metric $g_{\mu \nu}$. 
The interactions of the SM particles with the graviton, $G_{\mu\nu}$, 
can be derived from the following Lagrangian:
\begin{equation} 
{\cal L}=-{1 \over \bar M_P} G_{\mu \nu}^{(j)}T^{\mu\nu} ~,
\label{e2} 
\end{equation} 
where $j$ labels the Kaluza-Klein mode, $\bar M_P=M_P/\sqrt{8\pi}$
and $T^{\mu\nu}$ is the energy-momentum tensor. 
Given that the effective Lagrangian given in Eq.~\ref{e2} is suppressed
by $1/\bar M_P$, it may seem that the effects at colliders will be hopelessly
suppressed. However, in the case of real graviton production, the phase
space for the Kaluza-Klein modes cancels the dependence on $\bar M_P$ 
and, instead, provides a suppression of the order of $M_S$. For the
case of virtual production, we have to sum over the whole tower of 
Kaluza-Klein states and this sum when properly evaluated \cite{hlz, grw}
provides the correct order of suppression ($\sim M_S$). The summation
of time-like propagators and space-like propagators yield exactly the
same form for the leading terms in the expansion of the sum \cite{hlz}
and this shows that the low-energy effective theories for the $s$ and 
$t$-channels are equivalent.

Recently, several papers have explored the consequences of the above
effective Lagrangian for experimental observables at high-energy 
colliders. In particular, direct searches for graviton 
production at $e^+ e^-$, $p \bar p$ and $pp$ colliders, leading to 
spectacular single photon + missing energy or monojet + missing energy 
signatures, have been suggested \cite{grw, mpp, hlz, keung}. The virtual effects 
of graviton exchange in $e^+ e^- \rightarrow f \bar f$ and in high-mass 
dilepton production \cite{hewett}, in $t \bar t$ production \cite{us} 
at the Tevatron and the LHC, and in deep-inelastic scattering at HERA 
\cite{us2} have been studied. The bounds on $M_S$ obtained 
from direct searches depend on the number of extra dimensions. Non-observation
of the Kaluza-Klein modes yield bounds which are around 500 GeV to 1.2 TeV 
at LEP2 \cite{mpp, keung} and around 600 GeV to 750 GeV at Tevatron 
(for $n$ between 
2 and 6) \cite{mpp}. Indirect bounds from virtual graviton exchange
in dilepton production at Tevatron yields a bound of around 950 GeV 
\cite{hewett}. Virtual effects in $t \bar t$ production at Tevatron 
yields a bound of about 650 GeV \cite{us}, while from deep-inelastic
scattering a bound of 550 GeV results \cite{us2}. At LHC, it is expected
that $t \bar t$ production can be used to explore a range of $M_S$
values upto 4~TeV \cite{us}. 
More recently, these studies have been
extended to the case of $e^+ e^-$ and $\gamma \gamma$ collisions
at the NLC \cite{rizzo, soni}. There have also been papers discussing the
implications of the large dimensions for higgs
production \cite{rizzo2} and electroweak precision observables \cite{precision}. 
Astrophysical constraints, like bounds from energy loss for supernovae cores, 
have also been discussed \cite{astro}.

In the present work, we study the effect of the virtual graviton
exchange on the dijet production cross-section in $p \bar p$ collisions
at the Tevatron. 
The presence of the new couplings from the low-energy effective
theory of gravity leads to new diagrams for dijet production.
Using the couplings
given in Refs.~\cite{grw, hlz}, and summing over all the graviton modes,
we have calculated the sub-process cross-section due to the new
physics \footnote{The explicit expressions for the subprocess cross-sections
will appear in a future publication \cite{future}}. The graviton induced 
cross-sections 
involve two new parameters~: the effective string scale $M_S$ and
$\lambda$ which is the effective coupling at $M_S$. $\lambda$ is expected 
to be of ${\cal O}(1)$, but its sign is not known $a\ priori$.
In our work we will explore the sensitivity of our results to the 
choice of the sign of $\lambda$. 
\begin{figure}[ht]
\begin{center}
\vspace*{3.7in}
      \relax\noindent\hskip -4.7in\relax{\includegraphics{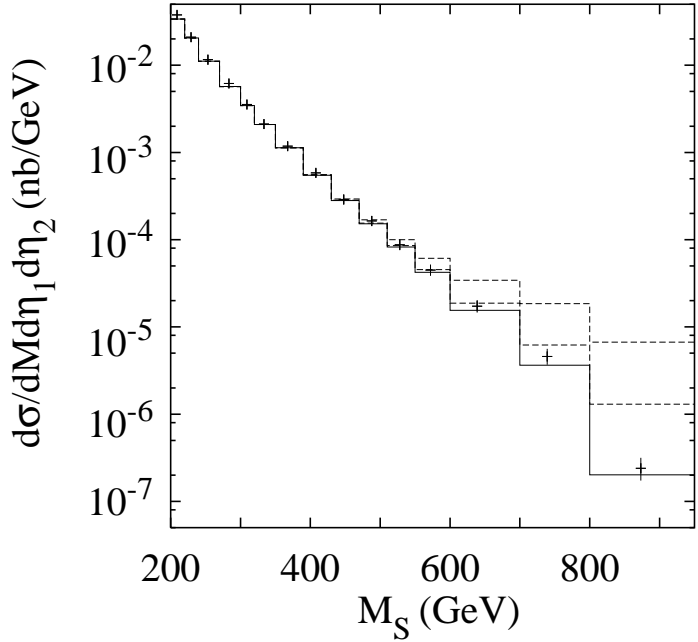}}
\end{center}
\vspace{-0.8in}
\caption{\footnotesize\it Illustrating the variation of the dijet
                          mass distribution with variation in the scale
                          $M_S$ at the Tevatron. The solid histogram
                          shows the SM NLO prediction; dashed histograms
                          show the prediction for $M_S=800$ GeV and 1 TeV
                          (upper line and lower line, respectively.) 
                           Data are taken from the D0 collaboration. } 
\end{figure}
%

Significant changes in the angular distribution of jets is expected when
spin-2 particle exchanges are added to the spin-1 exchange of the SM.
With this in mind, we study the normalised $\chi$ distribution, $1/N
dN/ d\chi$, where the variable $\chi$ is defined as 
\begin{equation} 
\chi={\hat u \over \hat t} \equiv {\rm exp} \vert \eta_1 - \eta_2 \vert,
\label{e3} 
\end{equation} 
with $\eta_1$ and $\eta_2$ being the pseudo-rapidities of the two jets,
so as to be able to compare with the experimental results from the 
CDF \cite{cdf} and the D0 \cite{d0} collaborations. 
The $\chi$ distributions in both the experiments have been calculated
in different mass bins, and we have used the same binning as used by
the two experiments. Using the same kinematic cuts as used
by the experimentalists (insofar as can be implemented in a parton-level
analysis), we study the normalised $\chi$ distribution
as a function of the effective string scale, $M_S$, and obtain the
95\% C.L. limits on the string scale by doing a $\chi^2$ fit to the
data in each bin and to the data integrated over the entire mass range. 
For our computations, we have used CTEQ4M parton densities \cite{cteq} 
taken from PDFLIB \cite{pdflib}. 
The 95\% C.L. limits on $M_S$ derived from the CDF and the D0 $\chi$
distributions, respectively, are displayed in Tables 1 and 2 for the cases
$\lambda= \pm 1$. We find that the $\chi$ distribution integrated over
the entire mass range yields a limit of 1070 (1108) GeV for $\lambda=1 (-1)$
for CDF and a limit of 1160 (1159) GeV for $\lambda=1 (-1)$ for D0. These
bounds are the most stringent bounds obtained from processes involving virtual
graviton exchange. Interestingly, the bounds obtained by considering the
highest mass bin are almost as good as those obtained by comparing with
all the data. This tells us that the deviations from the SM is greater as the
invariant mass increases. We, therefore, consider the data in the invariant
mass distribution as well.

\begin{table}[htbp]
\begin{center}
95\% C.L. limits on $M_S$ (in GeV) derived from the CDF $\chi$ distribution
\begin{tabular}{|c|c|c|c|c|c|c|}
\hline
Bin&241-300&300-400&400-517&517-625&$> 625$&Combined\\
\hline
+1&585&587&753&873&1095&1070\\
-1&626&544&717&852&1075&1108\\
\hline
\end{tabular}
\end{center}
\end{table}

\vskip-20pt
\begin{table}[htbp]
\begin{center}
95\% C.L. limits on $M_S$ (in GeV) derived from the D0 $\chi$ distribution
\begin{tabular}{|c|c|c|c|c|c|}
\hline
Bin&260-425&425-475&475-635&$> 635$&Combined\\
\hline
+1&523&632&919&1154&1160\\
-1&500&614&896&1131&1159\\
\hline
\end{tabular}
\end{center}
\end{table}
Recently, dijet mass distributions from the D0 experiment have
become available \cite{d02}. We have studied these 
(using the cuts used by the D0 experiment) and obtain,
as before, the 95\% C.L. limits on $M_S$. In Fig.~1, we have plotted
the mass distribution for different $M_S$ values and compared it
to the experimental and the SM numbers. We find again that very
stringent bounds for both signs of the $\lambda$ coupling are obtained.
For $\lambda=1$, we find that the 95 \% C.L. limit on $M_S$ is 1123 GeV,
whereas for $\lambda=-1$ it is 1131 GeV. Since the effect of the new physics
is larger for larger values of dijet mass, we find that if we use a
lower cut of 500~GeV on the dijet mass the resultant 
$\chi^2$ fit can yield a better limit on $M_S$. 

We have studied the implications of large extra dimensions and a low 
effective quantum gravity scale for dijet production at the Tevatron. Virtual
exchange of the Kaluza-Klein states are considered and the sensitivity
of the experimental cross-sections to this interesting new physics is
studied. We find that this process allows us to put very stringent limits
on the effective string scale $M_S$ -- in fact, of all processes with
virtual graviton exchanges considered so far, these bounds are by far
the best. To obtain these bounds, we have considered the angular 
distributions and the mass distributions. The resulting limits from
either of these observables are quite similar. Jet production at higher
energies is able to probe the physics of large extra dimensions
to much higher scales. These results will be presented in a future
publication \cite{future}.

\vskip20pt
{\sl Acknowledgements:} It is a pleasure to thank T.~Askawa for
help with the experimental data and Dilip K. Ghosh for discussions.

\end{document}